\documentclass[12pt]{iopart}
\usepackage{iopams}
\usepackage{setstack}

\usepackage[english]{babel}
\usepackage{cite}
\usepackage[pdftex]{graphicx}
\usepackage[pdftex]{hyperref}
\usepackage{lscape} 
\usepackage{pdflscape} 
\usepackage{rotating}
\usepackage[usenames]{color}
\usepackage[normalem]{ulem}
\usepackage{ulem} 
\usepackage{float}
\usepackage{subcaption}
\usepackage{amstext}
\usepackage{changepage}

\newcommand{\abs}[1]{\left| #1 \right|} 
\newcommand\bluesout{\bgroup\markoverwith{\textcolor{blue}{\rule[0.5ex]{4pt}{1.5pt}}}\ULon} 

\begin{document}

\title[]{Information-theoretic measures of electron correlation in a double quantum dot}

\author{W. S. Nascimento$^{1,\dagger}$, A. M. Maniero$^{2,\ddagger}$, F. V. Prudente$^{1,\S}$, Ginette Jalbert$^{3,**}$}

\address{$^1$Instituto de F\'{\i}sica, Universidade Federal da Bahia, 40170-115, Salvador, BA, Brazil. 

$^2$Centro das Ci\^encias Exatas e das Tecnologias, Universidade Federal do Oeste da Bahia, 47808-021, Barreiras, BA, Brazil. 

$^3$Instituto de F\'{\i}sica, Universidade Federal do Rio de Janeiro, 21941-972, Rio de Janeiro, RJ, Brazil.}

\ead{$^{\dagger}$wallassantos@gmail.com, $^{\ddagger}$angelo.maniero@ufob.edu.br, $^{\S}$prudente@ufba.br, $^{**}$ginette@if.ufrj.br.}

\vspace{10pt}


\begin{abstract}

In this work, we investigated how variations in spatial confinement affect the electron correlation between two electrons in a symmetric double quantum dot. In strong confinement regime, the ground-state energy can be obtained using the Hartree-Fock method neglecting the electron-electron repulsion term. Under a weaker confinement, the FCI method provides a more accurate description. Using the mutual information as a measure of correlation, we determine the confinement regions where electronic motion is uncorrelated. We conjecture that the Shannon entropy can serve as a qualitative guide to to study the electron correlation. We identified an approximately linear relationship between mutual information and relative correlation energy.

\end{abstract}

%
\vspace{1pc}

%
%
%
%

\section{Introduction}

Quantum dots are structures with dimensions on the order of the de Broglie wavelength formed at the interface between two semiconductors~\cite{Maniero_etal-2023,Maniero_etal-2021,Byakodi_etal2022,Das_Mandaletal-2025,Yakar_etal-2023}. These conductive nanostructures form a class of confined systems, where we can control, for example, their optical properties~\cite{Zhang_etal-2026,Kang_Wise-1997}. Implementing quantum bits, the quantum dots have been investigated in quantum computing~\cite{Jefferson_etal-2022}, and within the scope of quantum cryptography~\cite{Yang2024}, where they can function as single-photon emitters. 

The movement of the electrons in polyelectronic systems is correlated, since a change in the position of one electron alters the positions of the others. This electron correlation is due to the fermionic nature of electrons (Fermi correlation) and to Coulomb repulsion (Coulomb correlation)~\cite{Tew_etal2007,Sagar_etal-2011}. Fermi correlation acts on electrons with parallel spins and Coulomb correlation describes the repulsion between electron pairs independently of their spins~\cite{matta2024electron,Guevara_etal-2005}.

Electron correlation in quantum dots~\cite{Garcia_etal-2008,Kachu_etal-2025}, two-electron systems~\cite{Wilson_etal-2010,Yanajara-Parra_etal-2024}, and molecules~\cite{Fulde1995,Roztoczynska_etal-2025} has been studied in the literature. Determining electron correlation is essential for calculating dissociation energies~\cite{Martell_etal-1993}, in analysis of electron localization/delocalization~\cite{matta2024electron}, in studies on chemical shifts~\cite{Gauss-1993}, among other properties. Different theoretical measures based on energy have been proposed to analyze the phenomenon of correlation~\cite{correlacao_dif1, correlacao_dif3, correlacao_dif4, correlacao_dif5}; one of the most widely used is the so-called correlation energy, $E_{corr}$, defined as the difference between the exact and Hartree-Fock energies~\cite{Lowdin-1958}.

Since its first applications in physics, the Shannon information entropy has established itself as a tool for investigating quantum systems~\cite{Salazar_etal-2025,Kumar_Prasad-2025,Siddik_Sekh-2024,Arvizu_etal-2026}. 
From the perspective of quantum dots, information theoretic measures are important for mapping degenerate states and coupling between neighboring quantum dots~\cite{Nascimento_etal2024}, to indicate changes between symmetrical and asymmetrical regimes~\cite{Nascimento_etal2025}, to study the phenomenon of avoided crossing of the energy curves of different states~\cite{Guimaraes_etal2025}, to analyze electron entanglement~\cite{Mondal_etal-2022} and for the treatment of impurities~\cite{Liu_etal-2023}.   

Despite the success of energy analyses, several articles using informational entropies to study the electron correlation have been presented in the literature~\cite{Alipour2018,Kumar_Prasad-2026,nascimentocoulomb-et_al2021,Ramirez_etal1997,Esquivel_etal1996,Wang_etal2021,Guevara_etal-2003}. In this scenario, the mutual information, working with calculations that take as reference
separable pair densities,, has gained attention~\cite{Angulo_Rosa-2022,RPSagar_etal-2011,Kumar_Prasad-2023,Sagar_Guevara-2005,Zhao_etal-2025,SJCSalazar_etal-2026}. 
Furthermore, studies reveal how information entropy is related to fundamental physical quantities such as kinetic energy\cite{Massen_Panos-2001} and correlation energy~\cite{Collins1993}.

The aim of this study is to investigate the effects of electron correlation in a system in the ground state (singlet state) formed by two electrons confined in a double quantum dot. We are interested in analyzing how the relative correlation energy behaves when we alter the electronic confinement. We also investigate how  the Shannon entropy and mutual information can be used as a tool for studying electron correlation.

This article is organized as follows: section~\ref{theo_form} presents the theoretical framework; section~\ref{cal-met} presents the computational details of the calculations; section~\ref{analysis} discusses the results; and, section~\ref{conclusion} summarizes the central aspects of this investigation.

\section{Concepts}\label{theo_form}

In this section, we present the theoretical formalism of this work. Here, all physical formalism is developed in atomic units~(a.u.).

\subsection{System of interest}\label{Sys-in}

A system of two electrons confined in a symmetric double quantum dot (GaAs/AlGaAs heterostructure) can be described by the Hamiltonian
\begin{eqnarray}
    H =\sum_{i=1}^2 -\frac{1}{2m_c}\vec\nabla^2_i+ \sum_{i=1}^2 
        V(\vec r_i)
        +\frac{1}{\epsilon\vert\vec r_1-\vec r_2\vert} = T + V + V_{ee} \ , 
\label{hamiltoniano}
\end{eqnarray}
where the first term, $T$, is the electron kinetic energy term; the second term, $V$, is the potential energy term; and the third term, $V_{ee}$, is the electron-electron repulsion term. Besides, $\epsilon$ is the static dielectric constant and $m_c$ is the effective electronic mass. The confinement potential is given by~\cite{Nascimento_etal2024} 
\begin{eqnarray}
V(\vec r_i) = V(x_i,y_i,z_i) = {V}_{DQD}(x_i) + {V}_{HO}(y_i) + {V}_{HO}(z_i) \ ,
\label{potencialc}
\end{eqnarray}
given that
\begin{eqnarray}
{V}_{DQD}(x_i) =  V_0 \left[   A_0 \frac{x_i^2}{k^2} + B_0 e^{-\left(\frac{x_i}{k}\right)^2}  \right] \ , 
\  \ \ ( \textrm{with} \  A_0>0\ \ \textrm{and} \ B_0>A_0)\ , 
\label{HGA}
\end{eqnarray}
\begin{eqnarray}
{V}_{HO}(y_i) = \frac{1}{2}m_c \omega^2_y y_i^2 \ \ \ \ \ \text{and} \ \ \ \ \  {V}_{HO}(z_i) = \frac{1}{2}m_c \omega^2_z z_i^2 \ .
\label{HOYZ}
\end{eqnarray}
Here, ${V}_{DQD}(x_i)$ represents a symmetric double-well harmonic-Gaussian potential with $k$ relating the well to the barrier width (influencing the confinement condition), $V_0$ indicating the well depth, $A_0$ being associated with the well width, and $B_0$ corresponding to the inner barrier height. In turn, ${V}_{HO}(y_i)$ and ${V}_{HO}(z_i)$ are harmonic potentials where the frequencies $\omega_y$ and $\omega_z$ are confinement parameters. 

\begin{figure}[H]
\centering
\includegraphics[scale=0.26]{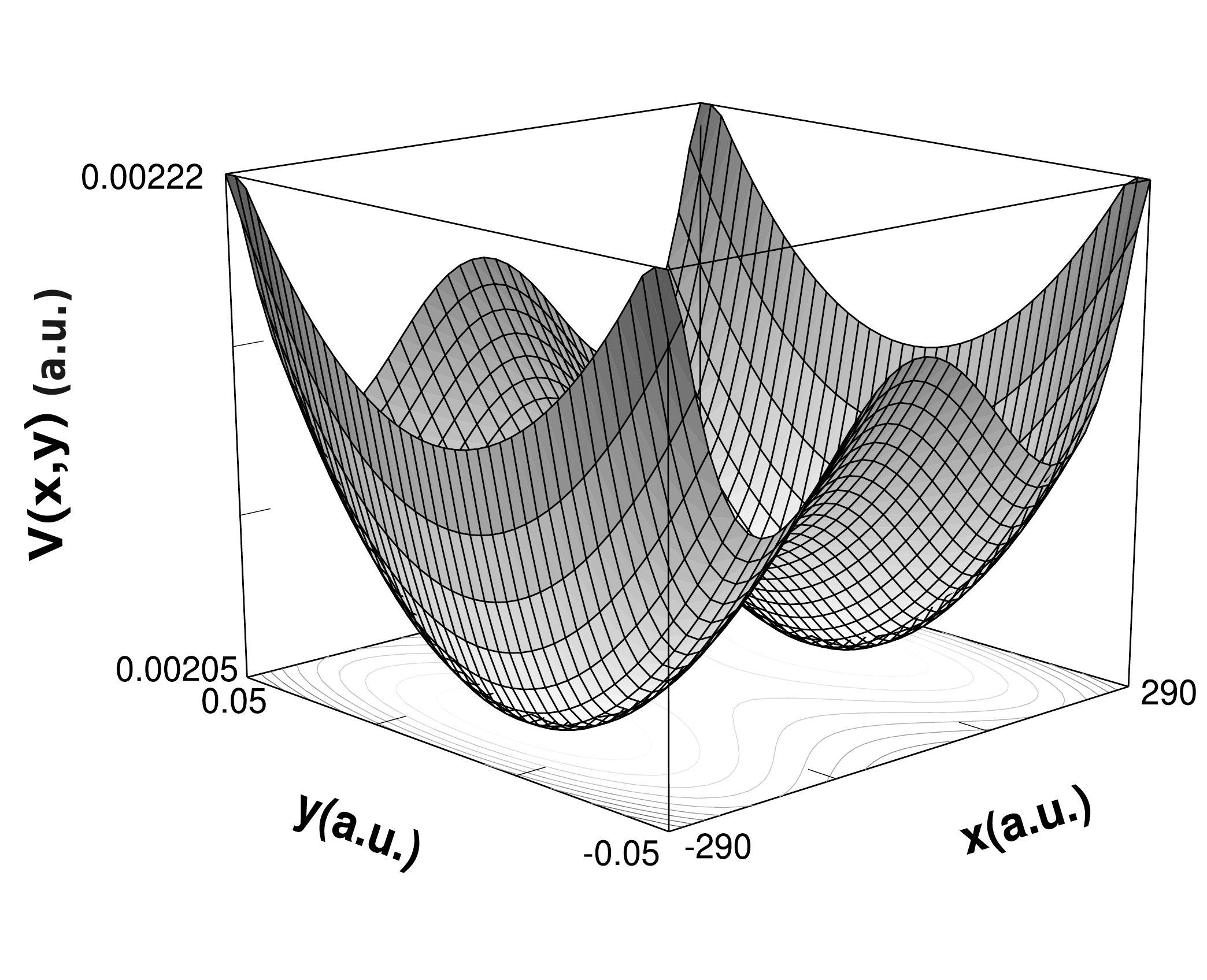}
\caption{Profiles of the confinement potential, $V(x, y, 0)$, for $A_0$~=~0.2, $B_0$~=~0.25, $k$~=~377.945~a.u., $V_0$~=~0.008379~a.u., $m_c$~=~0.067~a.u., $\epsilon$~=~13.1 and $\omega_y$~=~1.0~a.u..}
\label{Potencial3D}
\end{figure}

The quantum dots are centered on $X_\pm= \pm k\sqrt{\ln{\left({B_0}/{A_0}\right)}}$. In Fig. \ref{Potencial3D}, we present the profile of the confinement potential of the symmetric double quantum dot, from where we observe the symmetry of coupled quantum dots in respect to the internal barrier.

In our previous work involving one electron~\cite{Nascimento_etal2024}, we analyzed the symmetric double quantum dot through a detailed study of the confinement potential. In addition, we used informational quantities as a tool to investigate the degeneracy of quantum states and the levels of decoupling/coupling between coupled quantum dots.

\subsection{Hartree-Fock and Full Configuration Interaction Methods}\label{Metodos}

In Hartree-Fock (HF) method, the $N$-electron wavefunction, $\Phi_{\rm HF}$, is determined in terms of just one Slater determinant, such that~\cite{livrofred,szabo},   
\begin{equation}
\Phi_{\rm HF} = \frac{1}{\sqrt{N!}}|\chi_1(\vec{x}_1) \chi_2(\vec{x}_2)...\chi_N(\vec{x}_N) | \ ,
\label{funcaoHF}
\end{equation}
where, $\chi_i(\vec{x}_i)$ (with $i=1,2,..N$) are the spin-orbital functions written as the product of spatial functions, $\phi_i(\vec r)$, and spin functions, $\sigma_i(\omega)$. The $\phi_i(\vec r)$ are determined in such a way that
\begin{eqnarray}
\phi_i(\vec r)
=
\sum_\mu
C_{i \mu}
g_\mu(\vec r-\vec R_\mu;\vec{\zeta}_\mu).
\end{eqnarray}
Here, $C_{i \mu}$ are the coefficients of the expansion and $g_\mu(\vec r-\vec R_\mu;\vec{\zeta}_\mu)$ represent the basis functions centered at $\vec R_\mu$. From this scheme emerges the Hartree-Fock-Roothaan equation that, resolved by an auto-consistent process, provides the wavefunctions $\Phi_{\rm HF}$, and the Hartree-Fock energy, $E_{HF}$. 

The HF method assumes that each electron moves under the influence of the nuclei and of the average distribution of the other electrons. The Fermi correlation is introduced into the HF model by expressing $\Phi_{\rm HF}$ through a Slater determinant~\cite{matta2024electron,Tew_etal2007}. 

In Full Configuration Interaction (FCI) method, the $N$-electron wavefunction, $\Phi_{\rm FCI}$, is written as a linear combination of configuration state functions, $\Psi^{\rm{CSF}}$ (CSFs),~ therefore~\cite{livrofred,szabo},
\begin{eqnarray}
\Phi_{\rm FCI}=\sum_{i=1}^{N_{\rm{CSF}}}C_i^{\rm{CSF}}\Psi_i^{\rm{CSF}} \ ,
\label{funcaoFCI}
\end{eqnarray}
where $C_i^{\rm{CSF}}$ are expansion coefficients and $N_{\rm{CSF}}$ is the number of CSFs. Taking the Hartree-Fock function as reference, we have $\Psi_1^{\rm{CSF}} = \Phi_{\rm HF}$, and the CSFs being constituted by the sum of Slater determinants, i.e.,
\begin{eqnarray}
\Psi_i^{\rm{CSF}} = \sum^{Ndet_i}_{i_1}C^{det}_{i_1} det(i, i_1) \ ,
\end{eqnarray}
where $C^{det}_{i_1}$ are expansion coefficients and $det(i, i_1)$ is the $i_1$th determinant of the $i$th CSF. The FCI method involves solving an eigenvalue and eigenfunction problem from which we extract the wavefunctions, $\Phi_{\rm FCI}$, and the FCI energy, $E_{FCI}$.    

The FCI method provides a more precise calculation of the electronic interactions compared to the Hartree-Fock method. The Coulomb correlation, which is missing in the HF model, is introduced in the FCI model when $\Phi_{\rm FCI}$ is written as a linear combination of determinants~\cite{Tew_etal2007}. $E_{FCI}$ represents the exact solution within a given basis set.

\subsection{Informational quantities}{\label{information_entropies}}

In atomic and molecular physics, the Shannon entropy for a probability density in position space for an electron pair (Shannon pair entropies) is written as~\cite{nascimentocoulomb-et_al2021,Laguna_Sagar-2014} 
\begin{eqnarray}
S_{12} = -\int \int d\vec r_1 d\vec r_2 \rho(\vec r_1,\vec r_2)\ln[\rho(\vec r_1,\vec r_2)] \ .
\label{entropia2eletrons}
\end{eqnarray}
Here $\rho(\vec r_1,\vec r_2)= \vert\Phi(\vec r_1, \vec r_2)\vert^2 \ $, where $\vec r_1=(x_1,y_1,z_1)$ and $\vec r_2=(x_2,y_2,z_2)$ indicate the spatial coordinates of the particles. In turn, the entropy of the one-electron density is given by~\cite{Laguna_Sagar-2014}
\begin{eqnarray}
S_1 = -\int d\vec r_1 \rho_1(\vec r_1)\ln[\rho_1(\vec r_1)] \ ,
\label{entropia1eletrons}
\end{eqnarray}
with
\begin{eqnarray}
\rho_1(\vec r_1) = \int d\vec r_2\rho(\vec r_1,\vec r_2) \ .
\end{eqnarray}
The marginal density $\rho_1(\vec r_1)$ is normalized to unity. 
 
The entropy $S_r$ can be interpreted as a measure of delocalization of the probability distribution being physically dimensionless~\cite{nascimento-et_al2020,matta-etal2011}.

\subsection{Electron correlation measures}\label{ecm}

The correlation energy, $E_{corr}$, is defined as the error by which the Hartree-Fock energy, $E_{HF}$, differs from the energy $E_{FCI}$~\cite{Lowdin-1958}. However, $E_{corr}$ is not always a good measure of correlation~\cite{Ziesche_etal-1999}. Therefore, by considering more general cases, we adopt the relative correlation energy defined as
\begin{equation}
\beta = \abs{\frac{E_{FCI} - E_{HF}}{E^0}} \geqslant 0 \ ,
\label{enerfiacorrelacaor0}
\end{equation}
where $E^0$ is a reference energy. Specifically, for our system of interest, $E^0= E_{FCI} - \tilde{E}$, therefore, we can rewrite Eq.~(\ref{enerfiacorrelacaor0}) as
\begin{equation}
\beta = \abs{ \frac{E_{FCI} - E_{HF}}{E_{FCI}-\tilde{E}}} \geqslant 0 \ ,
\label{enerfiacorrelacaor}
\end{equation}
where $\tilde{E} = 2.004099$ a.u. is the total ground-state energy of the system in the asymptotic limit $k \rightarrow \infty$. We discuss the determination of $\tilde{E}$ in detail in Section 1 of the Supplementary Material.

In information theory context, we define mutual information as the amount of information shared between two variables~\cite{Cover,nielsen}. In atomic and molecular physics, the mutual information written as~\cite{Salazar_2025,Salazar_etal2-2025,Gutierrez_etal-2026} 
\begin{eqnarray}
I_r = \int \int d\vec r_1 d\vec r_2\rho(\vec r_1,\vec r_2)\ln\Bigg[\frac{\rho(\vec r_1,\vec r_2)}{\rho_1(\vec r_1)\rho_2(\vec r_2)}\Bigg]=2S_{1}-S_{12} \ \geq 0 \ ,
\label{entropiamutua}
\end{eqnarray}
and can be used as a measure of electron correlation. The product 
${\rho_1(\vec r_1)\rho_2(\vec r_2)}$ indicates that the probability of finding one electron at $\vec r_1$ is independent of the position of the other electron at $\vec r_2$, meaning that their spatial motions are uncorrelated. 

When defining the relative correlation energy, we adopt specific calculation methods, such as Hartree-Fock and FCI, and certain characteristics of the system; these dependencies limit the use of $\beta$ as a measure of correlation. In turn, the mutual information provides a more general measure of correlation, because it examines the statistical dependence between two variables. 

\section{Computational Details}\label{cal-met}

For the determination of the quantities of interest of this work, we consider $A_0$~=~0.2, $B_0$~=~0.25, $V_0$~=~0.008379~a.u., $m_c$~=~0.067~a.u., $\omega_y~=$~1.0~a.u.~and~$\omega_z$~=~1.0~a.u., $\epsilon$~=~13.1 and $\tilde{E}$~=~2.004099~\text{a.u.}.

We use anisotropic Gaussian functions as basis functions centered on $\vec R_\mu~=~(X_\mu,0,0)$, with $X_\mu~=~0,X_+,X_-$ and $n_{y_\mu}~=~n_{z_\mu}$~=~0, we write $g_\mu$ as 
\begin{eqnarray}
    g_\mu(\vec r-\vec R_\mu;\vec{\zeta}_\mu)=&&
    N_\mu
(x-X_\mu)^{n_{x_\mu}}
\times\nonumber\\
&&\exp\left[
-\zeta_{\mu x}(x-X_\mu)^2
-\zeta_{\mu y}y^2
-\zeta_{\mu z}z^2
\right].
\end{eqnarray}
In this scheme, we use 10 functions centered on the origin $X_\mu=0$, with functions 2s2p2d2f1g1h. The first exponent $\zeta_{\mu_x}^{(1)}$ is determined by minimizing the ground-state energy of a single electron in the confinement potential. We assume the same exponent for all functions $n_{\mu_x}$. The  second exponent is then obtained by multiplying $\zeta_{\mu_x}^{(1)}$ a scaling factor of 1.5. We use four more functions, two centered on $X_+$ and two centered on $X_-$, functions of the type s: with exponents 10 times and 25 times $\zeta_{\mu_x}^{(1)}$. Finally, we adopted the Gaussian exponents as $\zeta_{\mu_y} = \zeta_{\mu_z} = \frac{1}{2} m_c \omega$. Within the FCI method, the wavefunction contains 105 CSFs and 196 determinants. 

Throughout this work, for simplification, we will use the abbreviated notations $\rho(\vec r_1,\vec r_2)~=~\rho_{12}$, $\rho_1(\vec r_1)~=~\rho_{1}$ and $\rho_2(\vec r_2)~=~\rho_{2}$.

\section{Analysis and Discussion}\label{analysis}

Here, we discuss the values obtained for the quantities of interest of this work. In subsection~\ref{sobreadensidade}, we analyze the degree of delocalization 
of the probability density. In Subsection~\ref{sobreenergia}, we investigate the energies obtained using the Hartree-Fock and FCI methods, as well as the relative correlation energy. In Subsection~\ref{sobreanaliseinf}, we discuss mutual information as a measure of electron correlation and its behavior in the strong confinement regime. Finally, in Subsection~\ref{sobreinformacaoecorrelacao}, we present the relationships between information-theoretic quantities and correlation measures.

The calculations are performed in atomic units; when discussing confinement lengths, $k$ is reported in nanometer (nm), using 1~nm~$\approx$~18.897~a.u..

\subsection{Analysis of the Probability Density}\label{sobreadensidade}

\begin{figure}[H]
\centering
\includegraphics[scale=0.33]{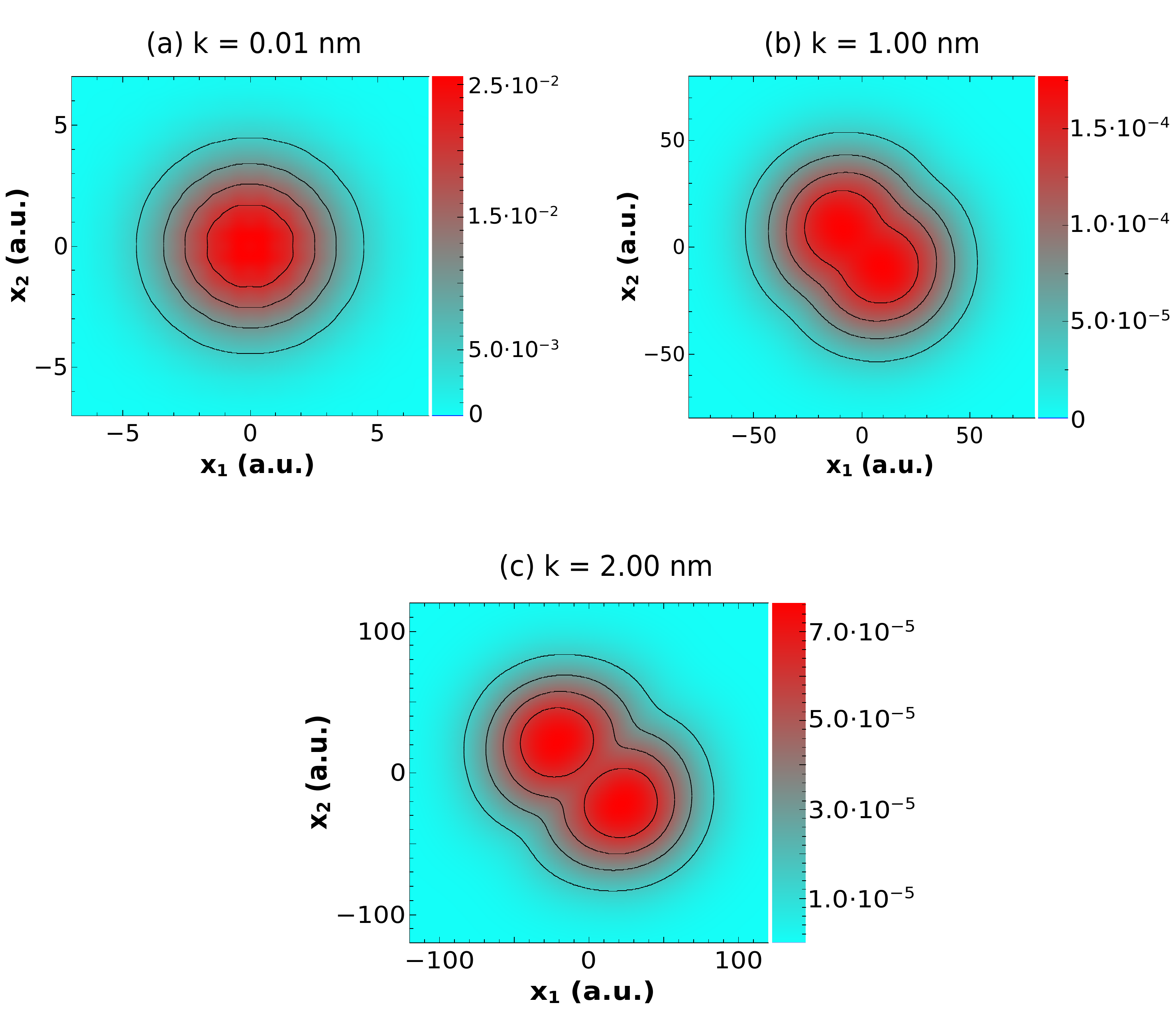}
\caption{Level curves of $\rho_{12} = (x_1, x_2)$ for the ground state of the system, obtained by the FCI method, for: (a) $k$~=~0.01~nm, (b) $k$~=~1.00~nm and (c) $k$~=~2.00~nm.}
\label{Densidade}
\end{figure}

In Figure~\ref{Densidade}, we present the level curves of $\rho_{12}$ for the ground state of the system, obtained via the FCI method, for different values of $k$. Specifically, in this analysis, we integrate the density over $y_1$, $y_2$, $z_1$ and $z_2$, in order to obtain $\rho_{12} = (x_1, x_2)$. In (a), with $k$~=~0.01~nm, the distance between the quantum dots is smaller than in cases (b) and (c); furthermore, $\rho_{12}$ reaches its lower degree of delocalization than in the other cases. In (b), with $k$~=~1.00~nm, the greater inter-dot separation leads to an increase in the degree of delocalization of $\rho_{12}$ compared to (a). In (c), with $k$~=~2.00~nm, $\rho_{12}$ exhibits a higher degree of delocalization than in (a)~and~(b).

According to Figure~\ref{Densidade}(a), the low degree of delocalization of $\rho_{12}$ is consistent with a stronger coupling between neighboring quantum dots. In contrast, in a configuration with weaker confinement, such as in Figure~\ref{Densidade}(c), the higher degree of delocalization of $\rho_{12}$ is associated with weaker coupling. In the asymptotic limit of $k\rightarrow \infty$, we infer that the system formed by two electrons in the ground state in a double quantum dot approaches the behavior of two decoupled systems, each consisting of a single electron in a dot.

\subsection{Analysis of the Energies}\label{sobreenergia}

\begin{figure}[H]
\centering
\includegraphics[scale=0.36]{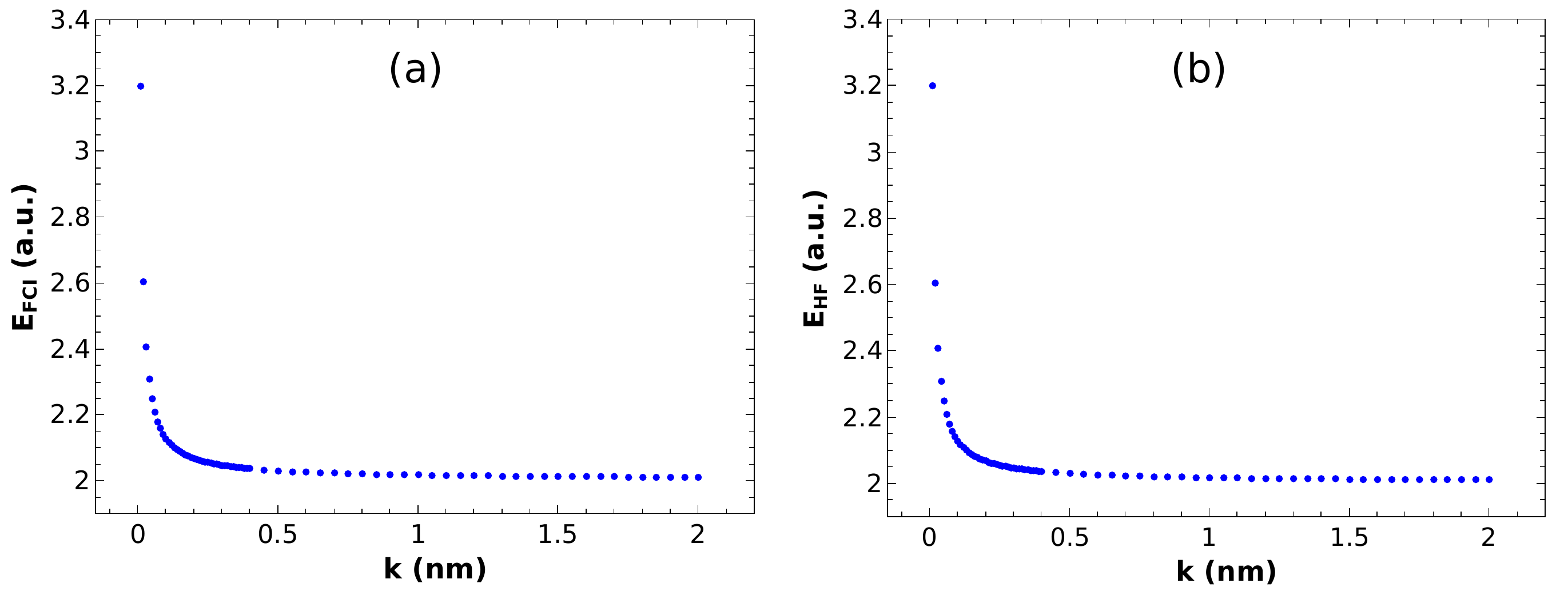}
\caption{In (a) energy $E_{FCI}$ and in (b) energy $E_{HF}$ as a function of $k$ for the ground state of the system.}
\label{EnergiaXk}
\end{figure}

We determined the energies $E_{FCI}$ and $E_{HF}$ for the ground state of the system. Figure~\ref{EnergiaXk}(a) shows the curve of the energy $E_{FCI}$ versus the parameter $k$. When $k \to 0$, the effects of electronic confinement increase,  resulting in a sharp increase in $E_{FCI}$. In contrast, as $k$ increases, the values of $E_{FCI}$ approach a limiting value. The curve of the energy $E_{HF}$ versus $k$, shown in Figure~\ref{EnergiaXk}(b), exhibits behavior similar to that observed for the curve in Figure~\ref{EnergiaXk}(a). We provide the values for the energies $E_{FCI}$ and $E_{HF}$ in Table S1 of the Supplementary Material.

\begin{figure}[H]
\centering
\includegraphics[scale=0.40]{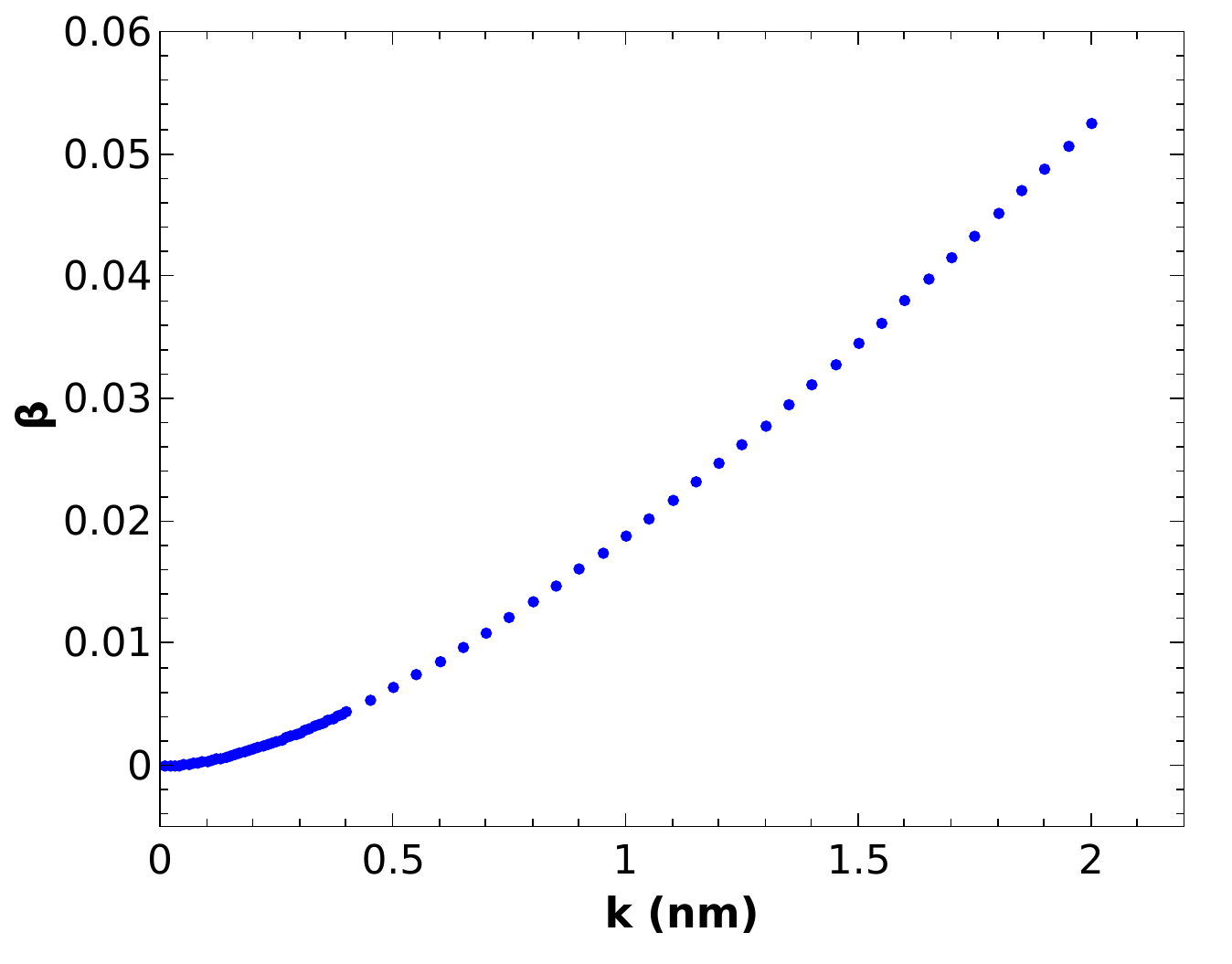}
\caption{Relative correlation energy, $\beta$, as a function of $k$ for the ground state of the system.}
\label{Gammaxk}
\end{figure}

Analyzing the curve of the relative correlation energy, $\beta$, versus $k$ for the ground state of the system, Figure~\ref{Gammaxk}, we observe that as $k$ decreases, the values of $\beta$ tend toward zero. Thus, in the asymptotic limit of $k \rightarrow 0$, corresponding to the regime of strong confinement, the total energy of the system can be successfully obtained using the Hartree-Fock method. When $k$ increases, the values of $\beta$ rise, indicating that the FCI method becomes necessary to accurately describe the total energy. We provide the values of $\beta$ in Table S1 of the supplementary material.

\begin{figure}[H]
\centering
\includegraphics[scale=0.40]{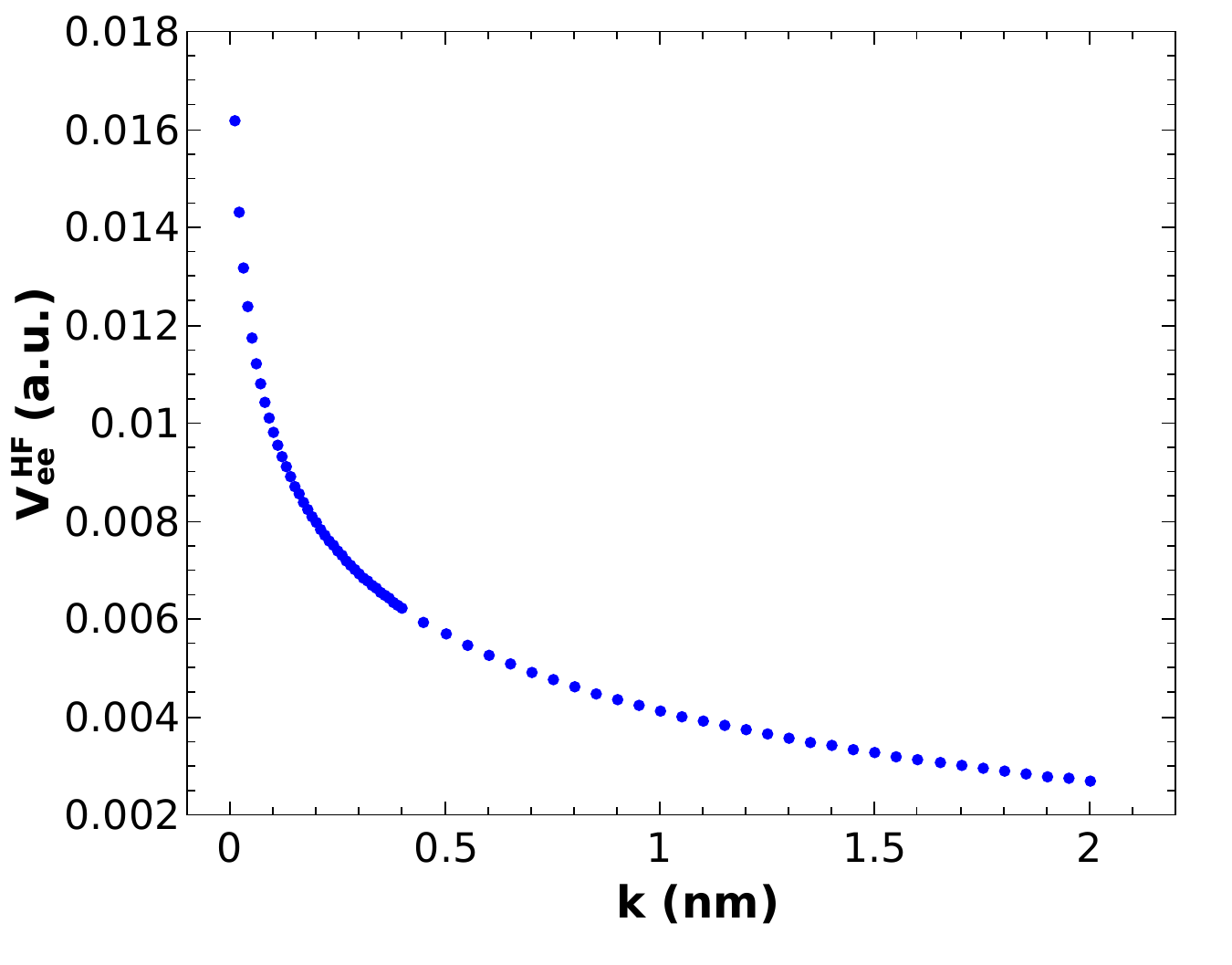}
\caption{Expectation value of the electron-electron repulsion energy obtained using the Hartree-Fock method, $V^{HF}_{ee}$, as a function of $k$ for the ground state of the system.}
\label{Veexk}
\end{figure}

Let us now examine the curve of the expectation value of the electron-repulsion energy obtained via the HF method, $V^{HF}_{ee}$, versus $k$ for the ground state of the system, as shown in Figure~\ref{Veexk}. As the values of $k$ decrease, the electron–electron distance tends to decrease, leading to an increase in $V^{HF}_{ee}$. Note that for small values of $k$, the kinetic energy component, $T^{HF}$, dominates the electron-electron repulsion energy, $V^{HF}_{ee}$; for instance, at $k$~=~0.01~nm, we have $T^{HF}~$=~1.590429~a.u. and $V^{HF}_{ee}$~=~0.016198~a.u. Thus, we conclude that in the asymptotic limit of $k \rightarrow 0$, the ground-state energy of the system can be successfully obtained using the Hartree-Fock method without explicitly accounting for the $V^{HF}_{ee}$ term.         
         
On the other hand, according to Figure~\ref{Veexk}, as the values of $k$ increase and the inter-dot distance grows, we observe a decrease in the electron-electron repulsion energy. In the asymptotic limit of $k\rightarrow \infty$, the term $V^{HF}_{ee}$ vanishes (see also Subsection 1.3 of the Supplementary Material), indicating a decoupling between the quantum dots, as discussed in subsection~\ref{sobreadensidade}.

\subsection{Analysis of Information-Theoretic Quantities}\label{sobreanaliseinf}

\begin{figure}[H]
\centering
\includegraphics[scale=0.40]{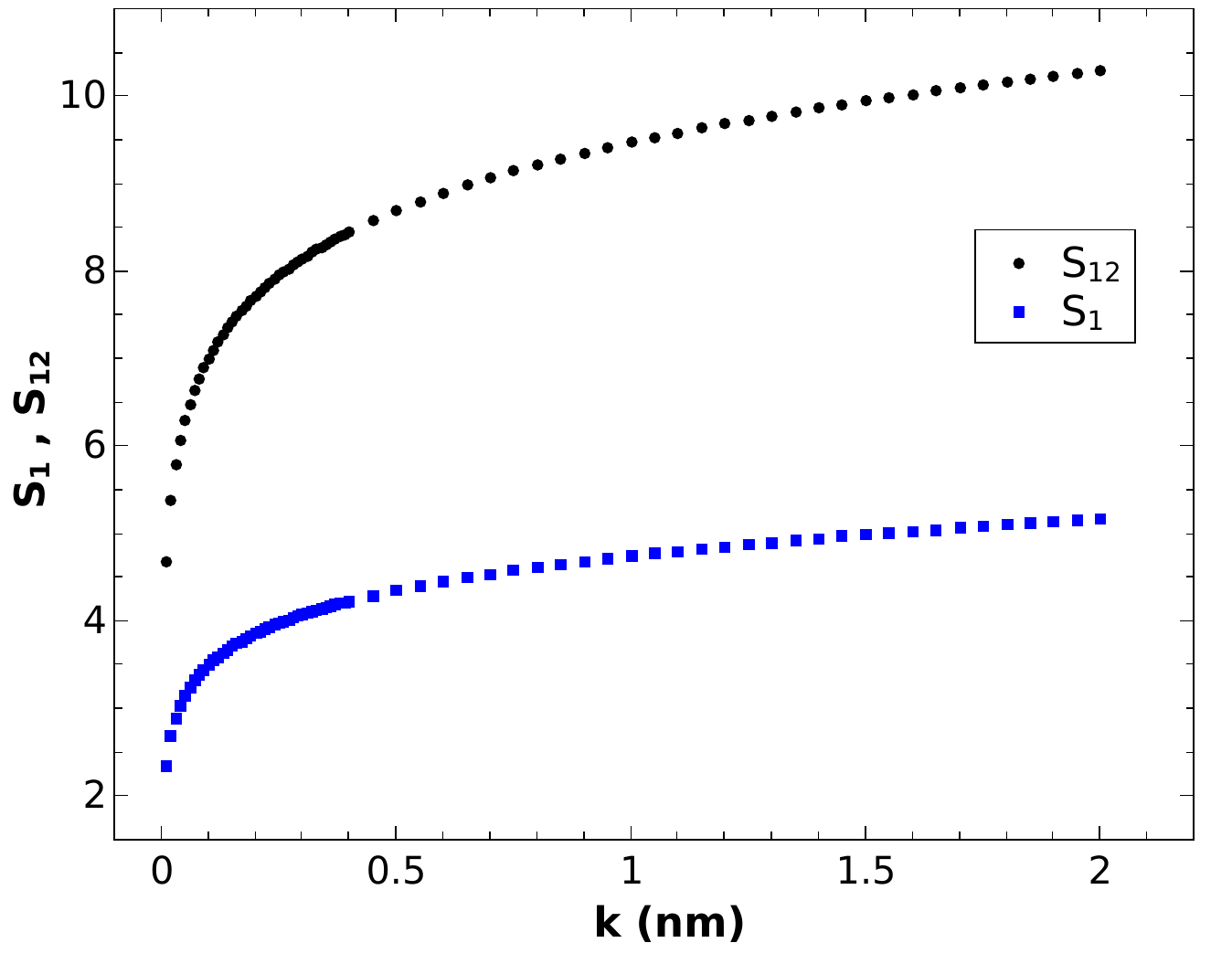}
\caption{Shannon entropy $S_{1}$ and $S_{12}$ as a function of $k$ for the ground state of the system.}
\label{entropias}
\end{figure}

Here, the values of $S_{1}$ and $S_{12}$ are calculated using the FCI method. Figure~\ref{entropias} shows the Shannon entropy curves $S_{1}$ and $S_{12}$ plotted as a function of $k$ for the ground state of the system. We first observe that the entropy values obtained for the electron-pair density are higher than those found for the one-electron density; that is, for a given value of $k$, the density $\rho_{12}$ exhibits a greater degree of delocalization than $\rho_1$. As confinement intensifies ($k \to 0$), the values of $S_{1}$ and $S_{12}$ decrease, indicating that $\rho_{1}$ and $\rho_{12}$ become less delocalized. Conversely, as $k$ increases, the values of $S_{1}$ and $S_{12}$ also increase, which is consistent with a greater degree of delocalization of $\rho_{1}$ and $\rho_{12}$. The values of $S_{1}$ and $S_{12}$ are provided in Table S2 of the Supplementary Material.

In summary, as the values of $k$ increase within the analyzed scope, $S_{12}$ also increases, implying a greater degree of delocalization of $\rho_{12}$. This behavior occurs in the region where electron correlation is relevant (see Subsection~\ref{sobreenergia}). Thus, we conclude that, in that case, for the ground state of the system, one of the consequences of correctly accounting for electron correlation is the emergence of more delocalized probability densities. Studies in the literature corroborate this result~\cite{shannon_correlacao2}.

\begin{figure}[H]
\centering
\includegraphics[scale=0.40]{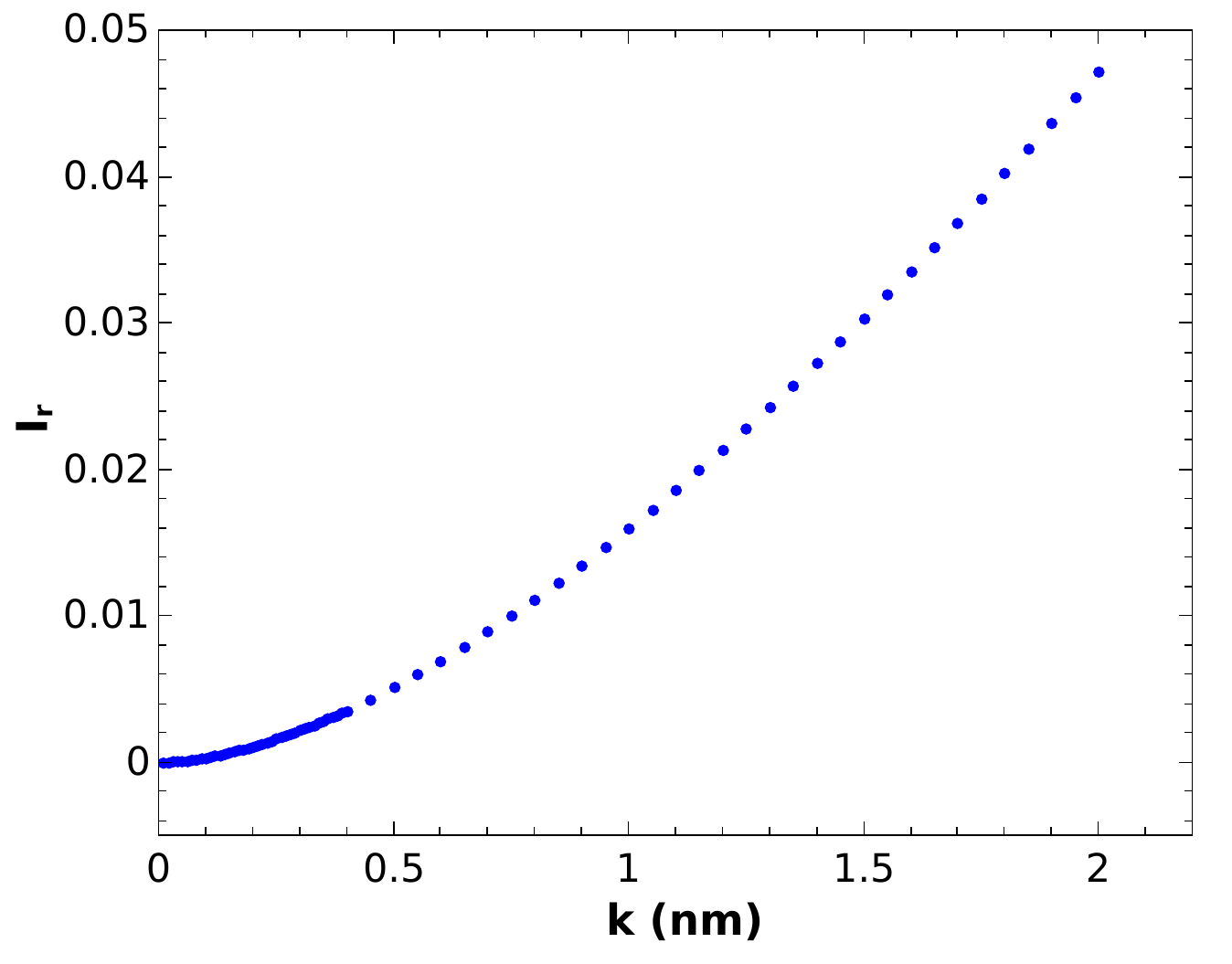}
\caption{Mutual information, $I_r$, as a function of $k$ for the ground state of the system.}
\label{Ixk}
\end{figure}

According to Figure~\ref{Ixk}, where we present the mutual information values, $I_r$, versus $k$ for the ground state of the system, the values of $I_r$ decrease and tend toward zero as $k$ decreases. (The values of $I_r$ are provided in Table S2 of the Supplementary Material). In other words, in the asymptotic limit $k\rightarrow 0$, the density $\rho_{12}$ can be decomposed as $\rho_1 \rho_2$, indicating that the spatial motion of the electrons is uncorrelated. As $k$ increases, the influence of electron correlation on spatial motion of the electrons becomes significant ($\rho_{12} \neq \rho_1 \rho_2$). Initial analyses noting that, in the strong confinement regime, the two-electron probability density can be decomposed into the product of the individual densities agree with our results~\cite{nascimentocoulomb-et_al2021,Santos_etal-2022}.

Note that the behavior of the plots in Figures \ref{Gammaxk} and \ref{Ixk} is quite similar, suggesting that mutual information can be successfully used as a measure of electron correlation in the context of this work.

\subsection{Relationships between Information-Theoretic Quantities and Correlation Measures}\label{sobreinformacaoecorrelacao}

\begin{figure}[H]
\centering
\includegraphics[scale=0.44]{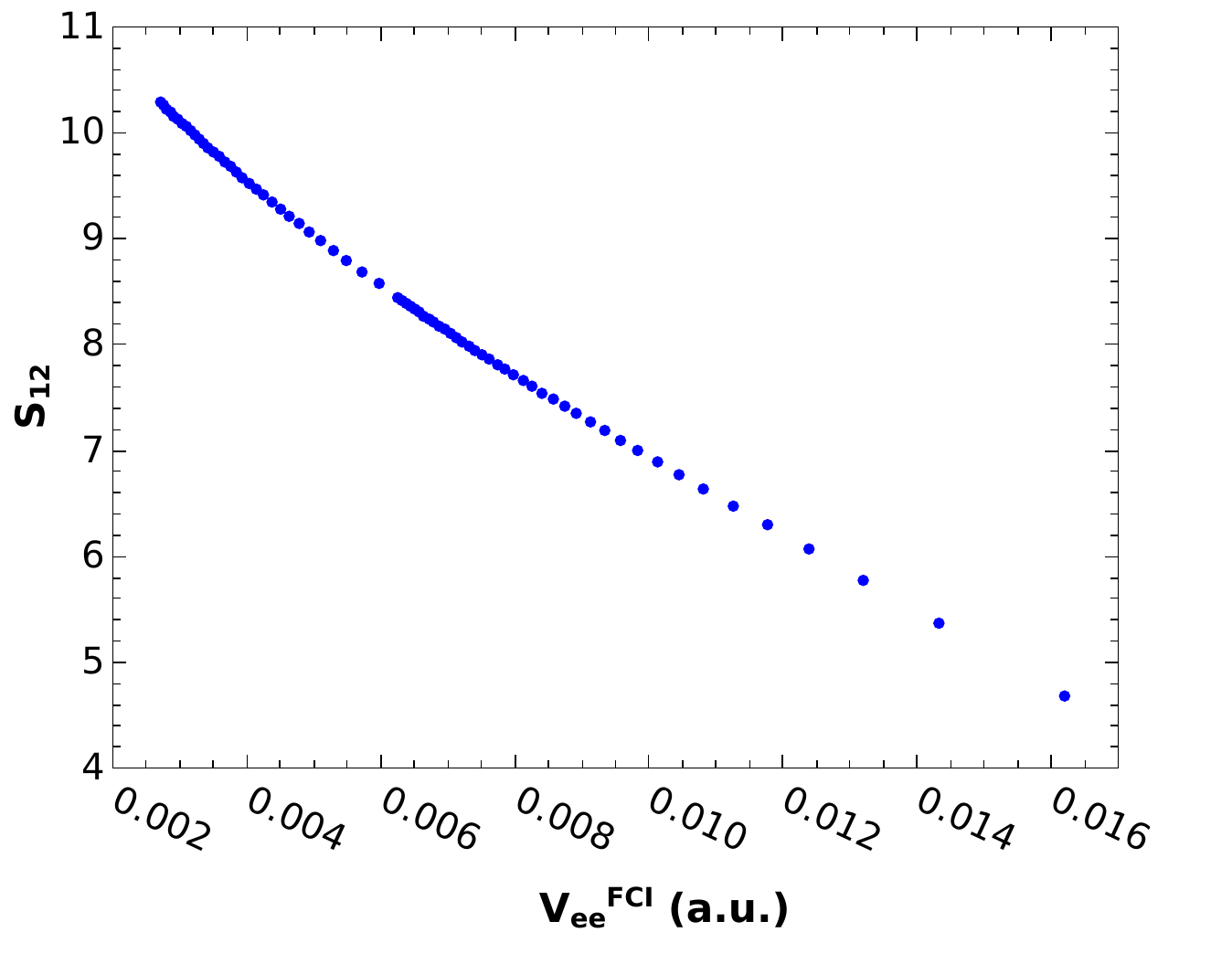}
\caption{Shannon entropy, $S_{12}$, as a function of the expectation value of the electron-repulsion energy, $V^{FCI}_{ee}$.}
\label{Sr1r2xVee}
\end{figure}

In Figure~\ref{Sr1r2xVee}, we present the values of the entropy $S_{12}$ versus $V^{FCI}_{ee}$, both quantities obtained for the ground state of the system using the FCI method. As $V^{FCI}_{ee}$ decreases, $S_{12}$ increases, and vice versa. In this way, low values of $S_{12}$ are associated with a higher degree of electron-electron repulsion; whereas high values of $S_{12}$ correspond to a lower degree of electron-electron repulsion. Thus, we suggest that the Shannon entropy, $S_{12}$, can serve as a qualitative guide for investigating electron-electron repulsion in this confined system.

\begin{figure}[H]
\centering
\includegraphics[scale=0.40]{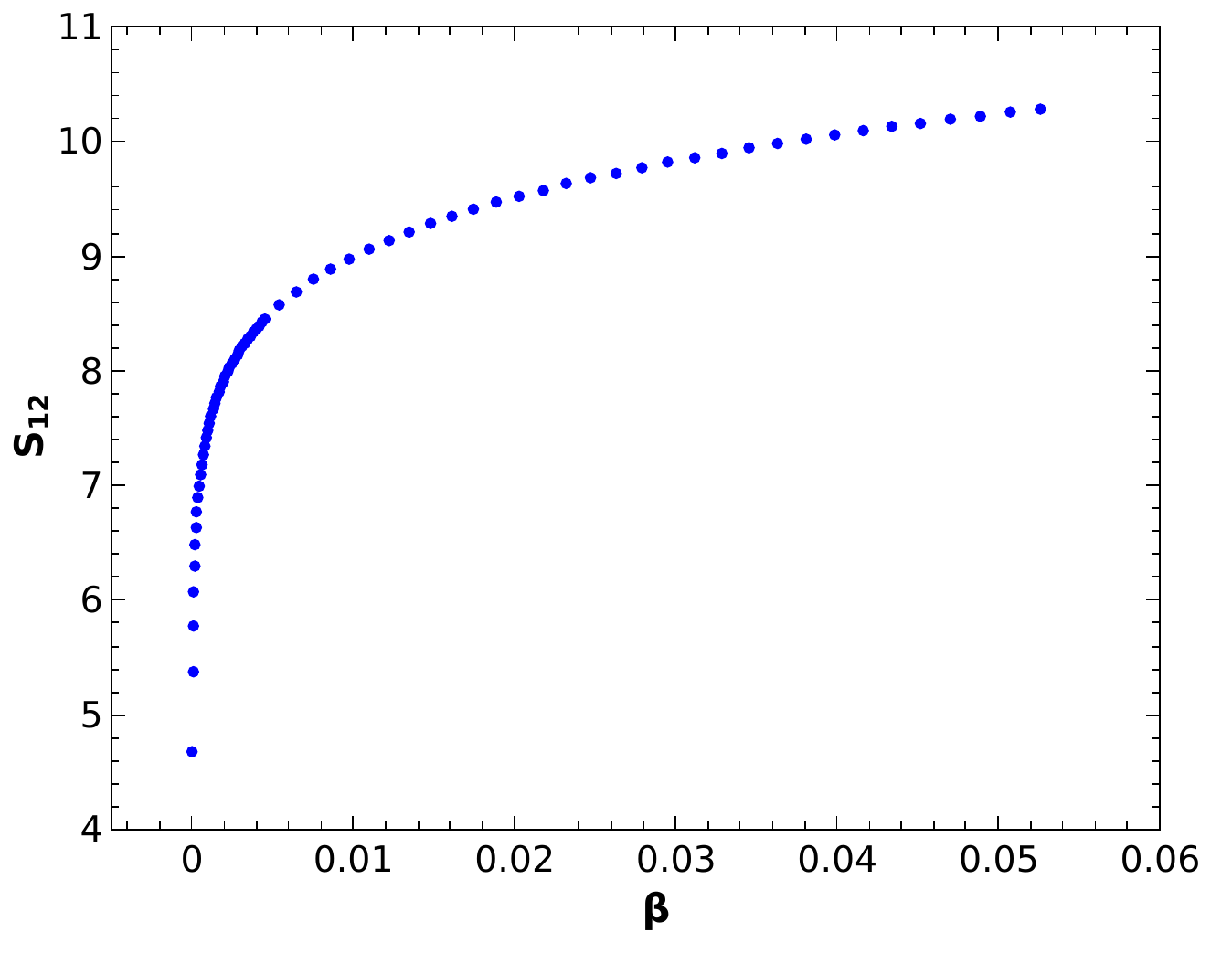}
\caption{Shannon entropy, $S_{12}$, as a function of the relative correlation energy, $\beta$.}
\label{Sr1Sr2xGama}
\end{figure}

In Figure~\ref{Sr1Sr2xGama}, we present the entropy $S_{12}$ versus the relative correlation energy, $\beta$ (ground state). We observe that as $\beta \rightarrow 0$, the values of $S_{12}$ decrease, indicating that $\rho_{12}$ tends to become less delocalized. As 
$\beta$ increases within the range investigated, the values of $S_{12}$ increase, reflecting an increase in the delocalization of $\rho_{12}$. Thus, in the system under study, the entropy $S_{12}$ can also be used as a qualitative indicator of electron correlation.

\begin{figure}[H]
\centering
\includegraphics[scale=0.40]{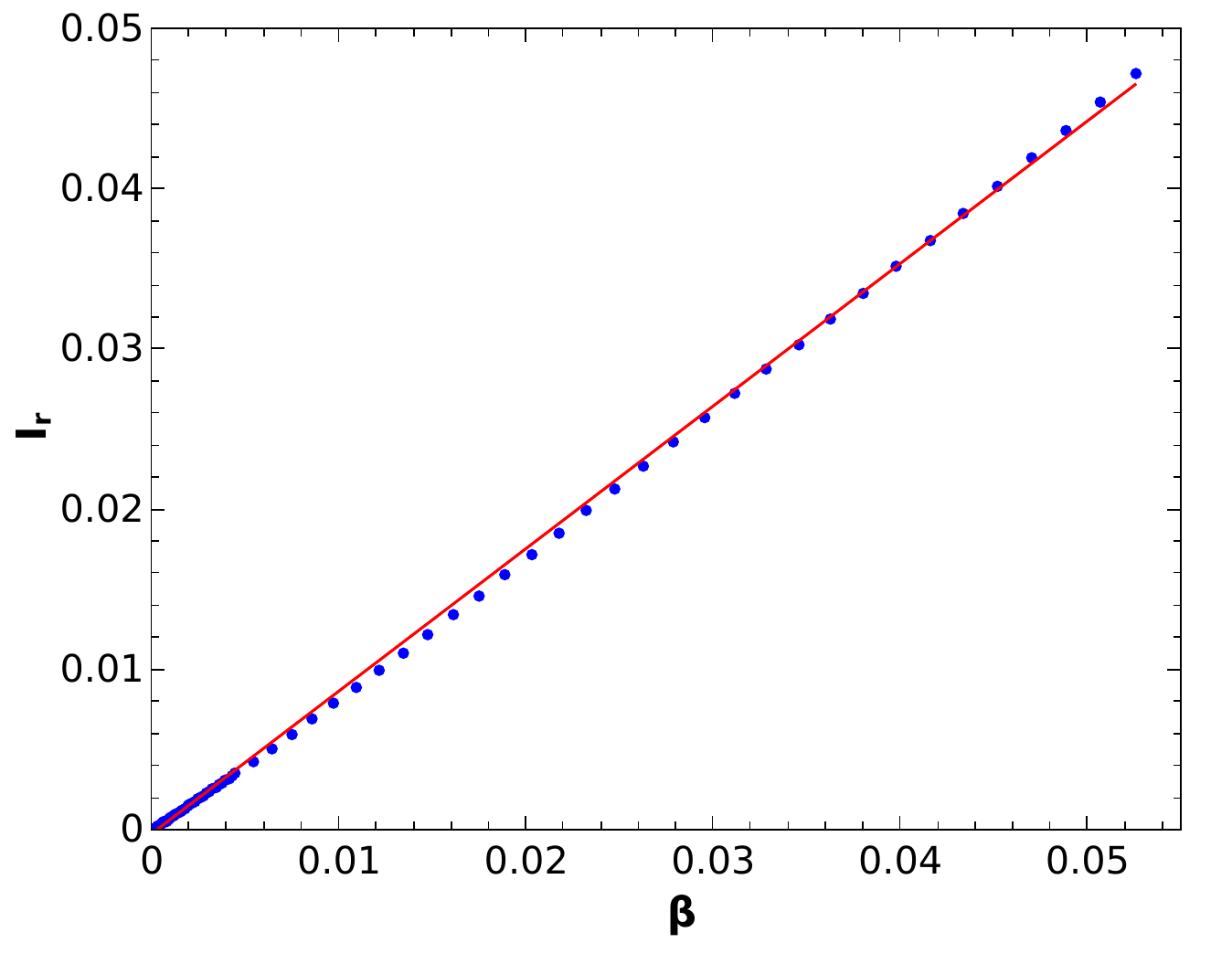}
\caption{Mutual information, $I_r$, versus relative correlation energy, $\beta$. The blue points correspond to the numerical values, while the red line corresponds to the linear fit~(\ref{ajustelinear}).} 
\label{Ixgamma}
\end{figure}

Despite the importance of the behaviors and trends established between $S_{12}$ and the energies $V_{ee}$ and $\beta$, we found an approximate match between the mutual information and the relative correlation energy for the ground state of the system. Figure~\ref{Ixgamma} shows the curve of $I_r$ versus $\beta$, where the blue points correspond to the numerical values we obtained. As $\beta$ increases, the quantity $I_r$ also grows, can be described by a linear fit of the form
\begin{eqnarray}
I_r = (0.890700 \pm 0.002304) \beta - (0.000309 \pm 0.000047) \ , 
\label{ajustelinear}
\end{eqnarray}
with R$^2$~=~0.999532. Thus, the red line in Figure~\ref{Ixgamma} explains approximately 99.9532\% of the variation observed in the data. Here, we identify an approximately linear relationship between $I_r$ and $\beta$, indicating that, within the model and confinement range investigated, $I_r$ can be successfully used as a measure of electron correlation. Previous studies point to technical advantages in adopting mutual information as a measure of correlation~\cite{Laguna_Sagar-2014}.

Historically, Collins is recognized as one of the pioneers in investigating and establishing direct relationships between informational and physical quantities. In particular, he demonstrated that the correlation energy is directly proportional to the Jaynes entropy~\cite{Collins1993}.

\section{Conclusion}\label{conclusion}

In a system formed by two electrons in a double quantum dot in the ground state, we investigated how the variation of spatial confinement impacts the electron correlation. To this end, we analyzed the energies obtained by the Hartree-Fock and FCI methods, and also the relative correlation energy and mutual information for the ground state of the system.

We find that in the situation where $k \to 0$, the degree of delocalization of $\rho_{12}$ decreases, indicating stronger coupling between neighboring quantum dots. Conversely, in the asymptotic limit $k \to \infty$, we infer that the system decouples into two independent systems, each consisting of an electron confined to a quantum dot.
 
We conclude that in the asymptotic limit of $k \rightarrow 0$, that is, in the regime of strong confinement, the ground-state energy can be successfully obtained using the Hartree-Fock method without explicitly accounting for the $V^{HF}_{ee}$ term. In the situation where $k \rightarrow \infty$, the FCI method provides a more accurate description. Furthermore, we find that one consequence of correctly accounting for electron correlation is the emergence of more delocalized probability densities. 

The mutual information analysis shows that, in the asymptotic limit $k\rightarrow 0$, $\rho_{12}~=~\rho_1 \rho_2$, indicating that the spatial motion of the electrons is uncorrelated. As $k$ increases, $\rho_{12}~\neq~\rho_1 \rho_2$, indicating the influence of spatial electron correlation. Furthermore, within the range of $k$ values analyzed, we conjecture that the Shannon entropy can serve as a qualitative guide to electronic correlation.

Finally, we established functional relationships between Shannon entropy and the electron-electron repulsion and relative correlation energies. We also identified an approximately linear relationship between mutual information and relative correlation energy. Such behaviors and relationships were established for the Hamiltonian parameters and confinement range considered here. In future work, we intend to investigate the fulfillment of these relations in other physical systems.

\vspace{2cm}

\noindent \textbf{Supporting Information}

This manuscript contains supplementary information.\\ 

\noindent \textbf{Acknowledgements}

The authors A.M.M., F.V.P. and G.J. acknowledge Conselho Nacional de Desenvolvimento Científico e Tecnológico (CNPq), Coordena\c{c}\~ao de Aperfei\c{c}oamento Pessoal de Nível Superior (CAPES) and Financiadora de Estudos e Projetos (FINEP) for the financial support. \\

\noindent \textbf{Conflict of Interest} 

The authors declare no conflict of interest.\\


\section*{References}

\bibliography{RefQDots8V2}
\bibliographystyle{unsrt}

\end{document}